# Machine learning regression analyses of intensity modulation two-photon spectroscopy (Mlim) in perovskite microcrystals

Qi Shi[1]*, and Tönu Pullerits [1]*

[1] The Division of Chemical Physics and NanoLund, Lund University, Box 124, 22100 Lund, Sweden

AUTHOR INFORMATION

**Corresponding Authors**

* qi.shi@chemphys.lu.se

* Tonu.Pullerits@chemphys.lu.se

**Abstract**

Perovskite thin films hold great promise for optoelectronic applications, such as solar cells and light emitting diodes (LEDs). A challenge is that defects are unavoidably formed in the material. Thorough understanding of the defect formation and their dynamics has proven challenging based on traditional spectroscopy. Here we integrated the functional intensity modulation two-photon spectroscopy with artificial intelligence (AI) - enhance data analyses to obtain a deep understanding of defect-related trap states within perovskite microcrystals.

We introduce a novel charge carrier recombination dynamics model that comprehensively includes exciton and electron-hole pair photoluminescence (PL) emissions, as well as the trapping and detrapping equilibrium dynamics. By varying parameters in the dynamic model, a large pool of the temperature dependent intensity modulation PL spectra can be simulated by solving the ordinary differential equations in the charge carrier dynamics model. Then, tree-based supervised machine learning methods (Decision tree, Extra tree, and Random Forest) and ensemble technique -- regression chain have been used to optimize the **M**achine **l**earning **i**ntensity **m**odulation spectroscopy (Mlim), which helps to determine the parameters of the charge carrier dynamics model (two-photon absorption, total trap concentration, trap activation energy) based on the temperature dependent intensity modulated PL spectra in perovskite. And the reliability of the Mlim predicted trap property parameters is confirmed by directly comparing the Mlim-retrieved intensity modulation spectra with experimental data.

Beyond Mlim's predictive capabilities, our approach unravels valuable insights into PL emissions, including those from excitons and free electron-hole pairs, but also provides details of trapping, detrapping, and nonradiative depopulation processes, offering a comprehensive understanding of perovskite material photophysics. This study suggests that Mlim applications hold promise for studying various photoactive devices.

## 1 Introduction

Hybrid perovskite semiconductors have recently garnered growing interest as materials for harvesting solar energy [1–3]. These materials exhibit exceptional optical properties, characterized by strong light absorption and efficient charge carrier generation, positioning them as significant contenders in the quest for sustainable energy solutions. However, the practical utilization of perovskite materials often faces challenges due to inherent imperfections, such as defects and trap states within their crystalline structures.

Understanding and characterizing properties of the traps within perovskite microcrystals are important tasks in optimizing their performance and stability of the material for the use in real-world applications. Intensity modulation two-photon excited PL microscopy (IM2PM) has emerged as a powerful technique that can provide detailed information about the dynamics of the traps and charge carriers in perovskite films and microcrystals[4–6].

Interpretation of the IM2PM signal is based on the analyses of the modulated emission in terms of Fourier components of the excitation modulation frequency. The model-dependent analysis of IM2PM data can be used to disentangle the PL emission using different charge carrier recombination orders[4,5]. Still, it is a challenge to separate and quantify the corresponding processes. For example, higher trap population, or smaller trap activation energy, both lead to more efficient first-order Shockley-Read-Hall (SRH) trap-assisted recombination – the dominant loss channel in perovskite-based photoactive devices. When the losses increase, what exactly is behind the increase? Clearly, a mechanistic understanding of the traps influencing the charge carrier recombination in the perovskite would be valuable. The charge carrier trapping and dynamics can be formulated as a set of coupled ordinary differential equations describing the details of the carrier recombination processes. The equations can be used to model experimental results allowing to evaluate important parameters responsible for recombination and other processes[7–10]. Since the number of variables can be large, finding a unique physically sound modelling solution that well describes IM2PM experiment, is very challenging.

Recently, machine learning (ML) methods have drawn significant attention in the scientific community including material science. ML has been applied for designing efficient functional molecules and materials[11–16]. Computational approaches like Monte Carlo, molecular dynamics, and density functional theory (DFT) augmented with ML methods are being used to identify high-performance photoactive perovskite materials, saving large amount of time-consuming and expensive laboratory screening work[17,18]. A deep neural network has been employed to address a constrained one-dimensional pulse retrieval challenge, demonstrating the capability for rapid, dependable, and comprehensive pulse characterization, which is achieved by leveraging interferometric correlation time traces generated from pulses that possess partial spectral overlap[19]. Combining large-scale simulation of the dynamics with the ML regression methods can pave the way to develop approaches that can accurately link the experimental observables like PL with the key processes in the perovskite thin films and gain a better understanding of the underlying photophysical phenomena.

In this work, we implement machine learning techniques to uncover the crucial charge carrier recombination processes affecting the Shockley-Read-Hall (SRH) losses in perovskite films based on the analyses of the IM2PM experiments. Initially, we perform extensive simulations of the experimental data varying a set of key model parameters that correspond to the methylammonium lead bromide ($MAPbBr_3$) perovskite film. Subsequently, we apply tree-based ML algorithms, and regression chain methods to the simulated data to verify the ability of the method to recover the input parameters. Among the models, the regression chain-extra tree method exhibits high accuracy in predicting the model parameters. The analyses allow to map out various charge carrier recombination processes, such as exciton PL emission, free electron-hole PL emission, trapping, detrapping, and SRH losses.

## 2 Data preparation for machine learning model training and validation

### Trap-mediated recombination model

We base our approach on a model which captures the physics of perovskite material and allows to analyze PL intensity as a function of time [7]. The following set of coupled ordinary differential equations (ODEs) (equations 1-3) describes trapping and dynamics of the photo-generated excitons, trap states and charge carriers:

$$\frac{dn_X}{dt} = G + R_f n_e n_h - R_d n_X - R_X n_X \,, \quad (1)$$

$$\frac{dn_e}{dt} = R_d n_X - R_f n_e n_h - \gamma_{trap} n_e (N_{Tr} - n_{Tr}) + \gamma_{detrap} n_{Tr} - R_{eh} n_e n_h \,, \quad (2)$$

$$\frac{dn_{Tr}}{dt} = \gamma_{trap} n_e (N_{Tr} - n_{Tr}) - \gamma_{detrap} n_{Tr} - \gamma_{depopu} n_{Tr} n_h \,. \quad (3)$$

The parameters $R_f$, $R_d$, $R_{eh}$, $R_X$, $\gamma_{trap}$, $\gamma_{detrap}$ and $\gamma_{depopu}$ are explained and descripted in Table 1. These parameters are well determined in earlier literature and are fixed accordingly. $G$ is the generation rate of excitons determined by the two-photon absorption. $n_X$ is the concentration of excitons. The activated electron trap concentration is a function of temperature, $N_{Tr} = N_{TR} * e^{-\frac{E_a}{k_B T}}$. $N_{TR}$ is the trap concentration including both active and passive traps. The trap activation energy $E_a$ is an important parameter that determines the likelihood of a trap being activated[20]. By trapping an electron, the trap becomes filled. $n_{Tr}$ is the concentration of the filled traps. We only consider electron trapping. Each trapped electron leaves behind a corresponding free photodoped hole in the valence band. Thus, the concentration of the free electrons at a certain temperature is $n = n_e$ and the concentration of the free holes is correspondingly $n_h = n_e + n_{Tr}$.

Table 1: Parameters used in the set of coupled ordinary differential equations for large scale data simulation. *RT* refers to room temperature.

| | Parameter | Symbol | Values | References |
|---|---|---|---|---|
| Fixed | Exciton formation rate from free electron and hole | $R_f$ | $10^{-12}\ cm^3 s^{-1}$ | 7 |
| | Exciton dissociation rate to electron and hole | $R_d(RT)$ | $5 * 10^{11}\ s^{-1}$ | 21 |
| | Trapping rate | $\gamma_{trap}$ | $8 * 10^{-9}\ cm^3 s^{-1}$ | 22 |
| | Detrapping rate | $\gamma_{detrap}(RT)$ | $10^{7}\ s^{-1}$ | 22 |
| | Electron hole radiative recombination rate | $R_{eh}$ | $5 * 10^{-11}\ cm^3 s^{-1}$ | 22 |
| | Exciton radiative recombination rate | $R_X$ | $5 * 10^{7}\ s^{-1}$ | 10 |
| | SRH nonradiative recombination rate | $\gamma_{depopu}(RT)$ | $5 * 10^{-9}\ cm^3 s^{-1}$ | 22 |
| Tunable | Generation rate of excitons | $G$ | $[10^{14}, 10^{15}]\ \ cm^{-3}$ | 5 |
| | Trap activation energy | $E_a$ | $[25, 200]\ (meV)$ | 23 |
| | Trap concentration | $N_{TR}$ | $[10^{21}, 10^{15}]\ cm^{-3}$ | |
| | Activated trap concentration | $N_{Tr}$ | $N_{TR} \times e^{-\frac{E_a}{k_B T}}$ | |

The exciton dissociation-association equilibrium is temperature dependent and follows the detailed balance relation as in the widely used Saha-Langmuir (S-L) equilibrium model[24,25],

$$\frac{R_d(T)}{R_f} = C * e^{-(\frac{E_b}{kT})}.$$

$E_b$ is the exciton binding energy of 84 meV [26]. At room temperature (*RT*), $C$ can be calculated using the values of $R_d(RT)$ and $R_f$. It is assumed that $R_f$ is less dependent on temperature. Subsequently, the temperature-dependent $R_d(T)$ can be calculated accordingly. Our model also includes charge carrier trapping-detrapping equilibrium, and the trapping rate is considered to be temperature independent. In the study by Herz et al. [22], the temperature-dependent detrapping rate is addressed through a thermally activated detrapping mechanism, encompassing changes in energy offsets and trap density near the phase transition temperatures. In the current study, we assume a thermally activated detrapping with the temperature dependent concentration of traps with certain activation energy. We do not consider changes to the energy offsets or trap density in the vicinity of phase transition temperatures.

Differently from most of the earlier studies we consider both exciton and free electron-hole pair radiative recombination as sources of photoluminescence (PL) emissions (PL $\sim R_X n_X + R_{eh} n_e n_h$). This is supported by the recent modeling of the absorption and emission of perovskite based on Elliott's theory which suggests that free electron-hole pairs do emit PL through radiative recombination[27]. Furthermore, Coulomb coupling between electron and hole can enhance the efficiency of this process [27,28]. Such studies highlight the importance of considering both exciton and free electron-hole pair contributions to PL emission in modeling and understanding of the optoelectronic device performance.

Auger recombination is not included in the model since the two-photon excitation concentration is low. The model also assumes that band-gap states are positioned above the Fermi level, and point defects primarily generate electron traps rather than hole traps in perovskite [22,29,30]. As a result, electron traps are assumed to be located energetically below the conduction band, while the conclusions for hole traps under analogous conditions are expected to be similar.

The irregularities in the micrometer-scale surface morphology of a perovskite polycrystalline film may significantly affect two-photon absorption ($G$). Additionally, the total trap concentration $N_{TR}$ as well as the trap activation energy $E_a$ play important role on affecting the non-uniform distribution of activated traps ($N_{Tr}$) within the film influencing PL intensity. These parameters are obtained from ML analyses of temperature dependent IM2PM experiments. We point out that only three parameters are free since $N_{Tr} = N_{TR} \times e^{-\frac{E_a}{k_B T}}$.

**Intensity modulation simulation**

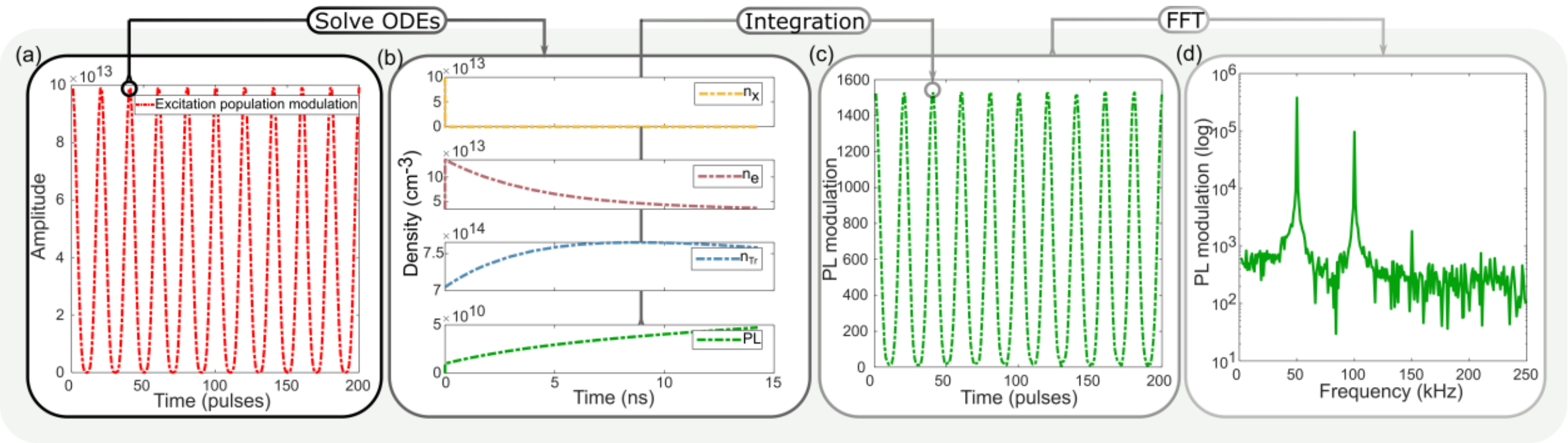

**Figure 1.** The workflow diagram of the excitation intensity modulation simulation. The time on X-axis in (a) and (c) is counted in the pulse intervals. (a) The red dash-dotted line provides the envelope of the modulated two-photon excited population. (b) the ODEs are numerically integrated by variable order method[31,32] giving the time dependent populations of the excitons (yellow), electrons (dark red) and filled traps (blue). The calculated populations are used to find the integrated PL intensity $\int(R_X n_X + R_{eh} n_e n_h)dt$ (green). (c) Modulated PL intensity. (d) frequency-domain signal is obtained by Fast Fourier transform (FFT) by MATLAB. A small noise was added in time domain to mimic the experimental conditions. The ODEs parameters used in this plot are shown in Table S1.

The two-photon excitation $G_i$ by i-th laser pulse depends on the laser pulse intensity square $\Psi$ (see Fig. 1a) and is given by equations 4-6 [5]

$$\Psi = 0.25 * (1 + \cos(\phi t))^2, \qquad (4)$$

$$\psi_i = \Psi(t_i), i = 1, \dots 2000, t_i \in [0\ s, 0.002\ s]\ , (5)$$

$$G_i = G * \psi_i, i = 1, \dots 2000\ .\ (6)$$

Here $\phi$ is the intensity modulation frequency. In the real experiment around 1400 pulses excite the sample during a single intensity modulation period. In the illustration for clarity, only 20 pulses are used during one modulation period, and a total of 1000 pulses ( $i = 1001 - 2000$) are used to simulate the experimental conditions within 1 millisecond. The initial 1000 pulses generate stable conditions and are not used for calculating the signal.

While solving ODEs for the Figure 1b, the inclusion of charge accumulation effect is essential for an accurate description of charge carrier recombination dynamics. Prior to the arrival of a new pulse $G_{i+1}$, a substantial fraction of charge carriers generated by the previous pulse are still present, in particular the trapped carriers can recombine very slowly. We obtain $n_X(t), n_e(t)$ and $n_{Tr}(t)$ from the rate equations (1-3) providing the initial conditions $n_X(t = 0s) = n_{X,start}, n_e(t = 0s) = n_{e,start}$ , and $n_{Tr}(t = 0s) = n_{Tr,start}$. $t \in \left[0\ ns, \frac{1}{f_{rep}}\right]$, with the laser repetition rate $f_{rep} = 70.14\ MHz$. For the first pulse excitation, $G_{i=1}$, $n_{X,i=1,start}$ $n_{e,i=1,start} = n_{Tr,i=1,start} = 0\ cm^{-3}$. The ODEs are numerically integrated by variable order method[31,32] , giving the time dependent populations of the excitons, electrons and the filled traps after an excitation pulse. Using the populations, the integrated PL $\int (R_X n_X + R_{eh} n_e n_h) dt$ is calculated. For the next pulse $G_{i=2}$, the starting charge carrier concentration is the accumulated charge carrier concentration at the end of last laser interval time period, $n_{X,i=2,start} = n_{X,i=1,end}$, $n_{e,i=2,start} = n_{e,i=1,end}$, and $n_{Tr,i=2,start} = n_{Tr,i=1,end}$.

Thereby, the starting values for solving the ODEs in each iteration are updated as shown in equations 7-9:

$$n_{X,i,start} = n_{X,i-1,end}, i \in [1001, 2000]\ (7)$$

$$n_{e,i,start} = n_{e,i-1,end}\ , i \in [1001, 2000]\ (8)$$

$$n_{Tr,i,start} = n_{Tr,i-1,end}\ , i \in [1001, 2000]\ (9)$$

All iterations are calculated using the same parameters.

The modulated PL emission is shown in Figure 1c. The Fast Fourier transform (FFT) via MATLAB is used to obtain the frequency-domain signal. The 50 kHz signal in figure 1d is called the basic harmonic or the first harmonic (A1H). The other three harmonics at 100, 150, and 200 kHz we call correspondingly A2H, A3H, and A4H.

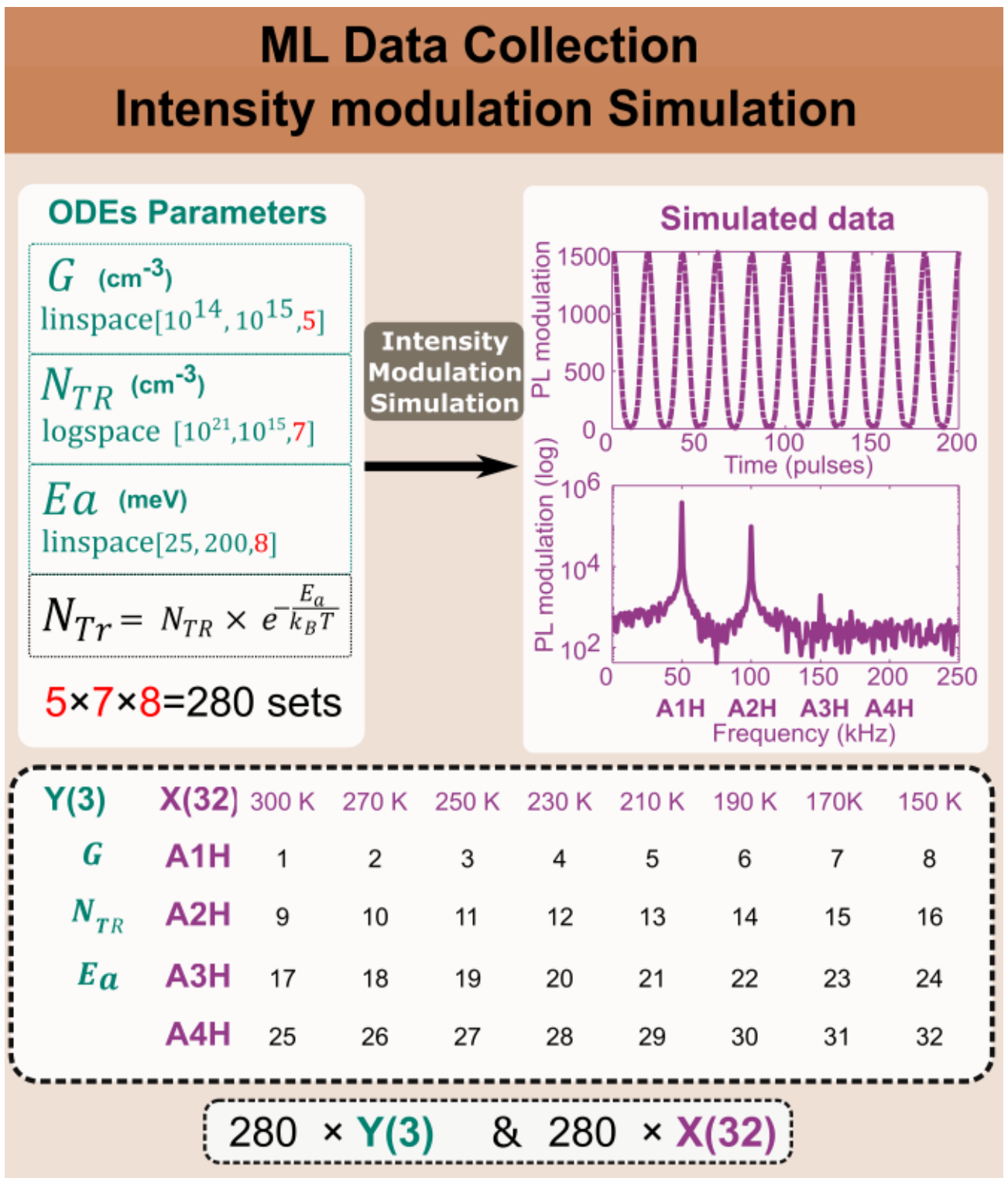

**Figure 2.** an overview of the process for intensity modulation simulation.

Based on the intensity modulation simulation procedure (Figure 1), for every set of the three parameters, we obtain 32 simulated data points $AiH(T)_{i=1...4}^{T=300...150\,K}$, consisting of amplitudes of the four harmonics (A1H, A2H, A3H, A4H) at 8 different temperatures (300, 270, 250, 230, 210, 190, 170, 150 K), shown in Figure 2. A total of 5 evenly spaced generation of exciton ($G$) values and 8 evenly spaced trap activation energy ($E_a$) values were generated within the specified ranges shown in Table 1. Similarly, 7 logarithmically spaced values for total trap concentration ($N_{TR}$) were generated. As a result, a total of 280 sets (5×8×7=280) of parameters ($G$, $E_a$, $N_{TR}$) for the ODEs, denoted as $280 \times Y(3)$, and the corresponding simulated intensity modulation spectra, denoted as $280 \times X(32)$, were generated.

The simulations mimic experimental data and offer insights into the photophysical properties of perovskite films under different conditions. Still, the reverse exercise, to obtain the parameters that describe the material from experimental results, is not trivial. Here we will use supervised machine learning techniques to address this issue.

## 3 Supervised machine learning regression architecture and training

Supervised machine learning regression is a branch of artificial intelligence. In this work we apply the method to map the simulated intensity modulation spectra, $280 \times X(32)$ into three charge carrier recombination dynamics parameters for the ODEs domain $280 \times Y(3)$ and make predictions for these parameters for the unseen data and for the future new experimental data. The detailed workflow of the training, validation, testing, and evaluation procedures of the supervised ML regressors is shown in Figure 3.

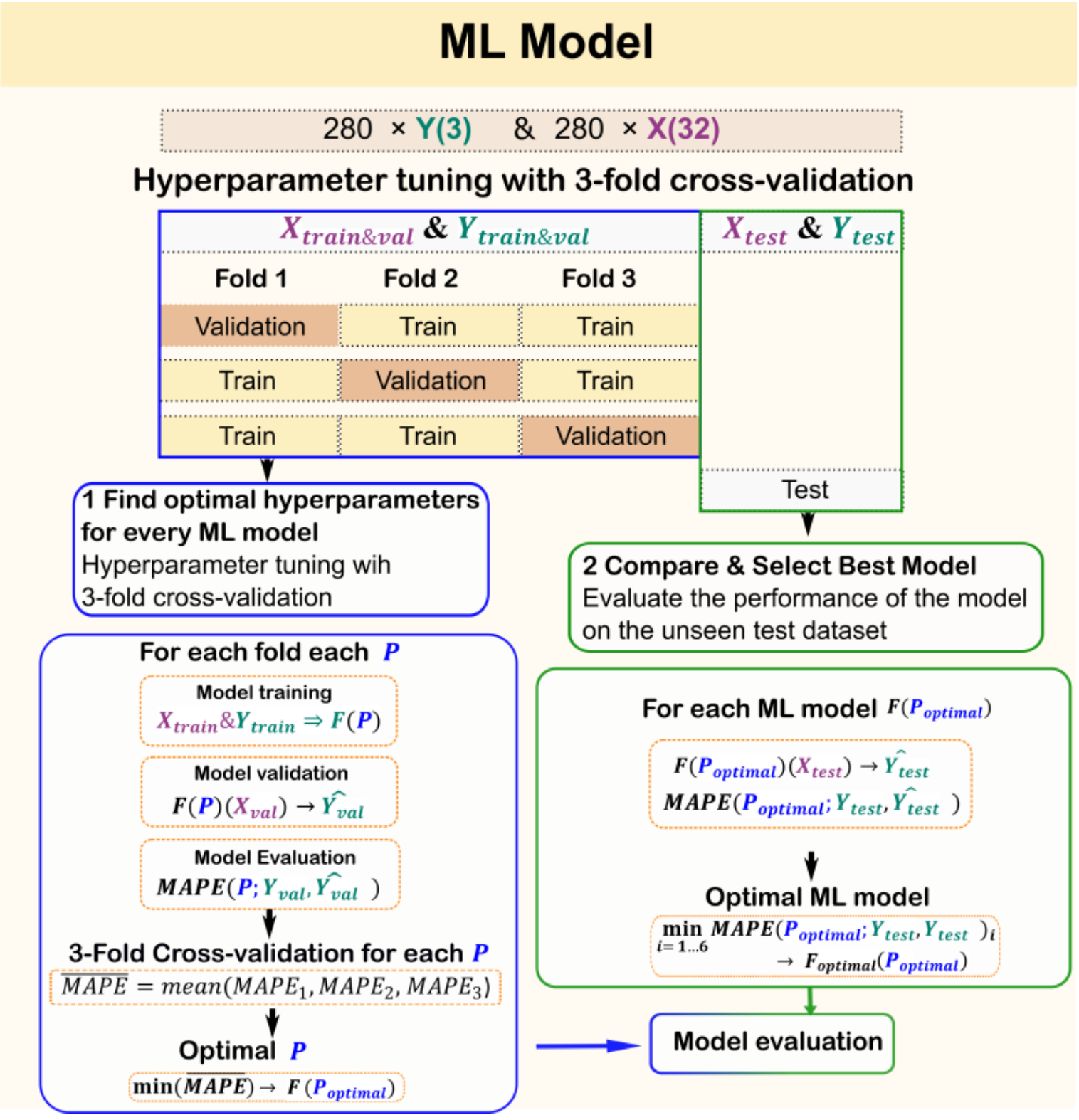

**Figure 3.** A flowchart of the study process of ML model.

To ensure that the machine learning models can predict well over a range of parameters, normalization methods are applied across the original simulated spectra data ($280 \times X(32)$). The normalized simulated dataset is first divided into a train-validation set ($\boldsymbol{X_{train\&val}}$ **&** $\boldsymbol{Y_{train\&val}}$, 90%, 252 sets) and a test set ($\boldsymbol{X_{test}}$ **&** $\boldsymbol{Y_{test}}$, 10%, 28 sets), marked with blue and green rectangles, respectively, in Figure 3. The 3-fold cross validation on train-validation dataset is used to optimize the so called 'hyperparameters (P)' of each ML model, which helps minimize prediction errors. Mean absolute percentage error (MAPE) is used as loss function here. And the test dataset is used as future unseen to evaluate the performance of each retrained ML model.

For each hyperparameter $P$, the model is trained using each fold of the training-validation dataset.

$$\boldsymbol{X_{train}}\&\boldsymbol{Y_{train}} \Rightarrow \boldsymbol{F}(\boldsymbol{P})$$

Then, the model is validated, and the loss was calculated using the Mean Absolute Percentage Error (MAPE) loss function between the prediction $\widehat{\boldsymbol{Y_{val}}}$ and the true label $\boldsymbol{Y_{val}}$ of the validation dataset.

$$\boldsymbol{F}(\boldsymbol{P})(\boldsymbol{X_{val}}) \rightarrow \widehat{\boldsymbol{Y_{val}}}$$

$$\boldsymbol{MAPE}(\boldsymbol{P}; \boldsymbol{Y_{val}}, \widehat{\boldsymbol{Y_{val}}})$$

After performing the same training-validation procedure with the remaining two folds, three losses ($\boldsymbol{MAPE}_1, \boldsymbol{MAPE}_2, \boldsymbol{MAPE}_3$) were obtained. An averaged loss is calculated to evaluate the performance of the current hyperparameter setting $P$.

$$\overline{\boldsymbol{MAPE}} = \text{mean}\ (\boldsymbol{MAPE}_1, \boldsymbol{MAPE}_2, \boldsymbol{MAPE}_3)$$

By exhaustively searching the entire combination of hyperparameters using grid search, the minimal loss is obtained along with the corresponding optimal hyperparameter values.

$$\min(\overline{MAPE}) \rightarrow F(P_{optimal})$$

Once each model is trained and validated with the optimal hyperparameters, it is ready for prediction. The unseen data $X_{test}$ is input into the trained model to obtain predictions.

$$F(P_{optimal})(X_{test}) \rightarrow \widehat{Y_{test}}$$

The loss is then calculated using the MAPE loss function between the predictions $\widehat{Y_{test}}$ and the true labels $Y_{test}$ of the unseen data.

$$MAPE(P_{optimal}; Y_{test}, \widehat{Y_{test}})$$

In this study, six machine learning regressors (decision tree, random forest, extra tree, regression chain-decision tree, regression chain-random forest, regression chain-extra tree) were selected to optimize performance. The minimal loss along with the corresponding different algorithms was assessed.

$$\min_{i=1\ldots6} MAPE(P_{optimal}; Y_{test}, \widehat{Y_{test}})_i \rightarrow F_{optimal}(P_{optimal})$$

The optimized hyperparameters ($P$) were described in S2.

**ML algorithms selection**

Decision tree (DT) is an efficient supervised learning algorithm which can be used for classification and regression[33,34]. A DT regressor is a hierarchical structure, and it predicts the values of a set of targe variables (a set of ODE parameters $280 \times Y(3)$) by learning if-then-else decision rules inferred from splitting the feature ranges $280 \times X(32)$ into small subdomains. A simple representation of decision tree for prediction of the one of the ODE parameters -- total trap concentration ($N_{TR}i$ ($i = 1 \ldots 6$)) using the spectra features at room temperature ($AiH(T)_{i=1\ldots4}^{T=298\,K}$), is shown in S3. It is transparent to interpretation and easy to visualize but can be unstable to variations in the data and is prone to over-fitting.

Random Forest (RF) Regression[35–37] and Extremely Randomized Trees (Extra Trees/ET) Regression[33,38] are both ensemble methods that use multiple decision trees for regression problems. RF creates multiple decision trees and averages their predictions to reduce the variance and overfitting. The ET algorithm takes it a step further by randomly selecting splits for the features ($AiH(T)_{i=1\ldots4}^{T=298\,K}$), which can lead to increased diversity and better performance in certain cases. Overall, these algorithms aim to find the best split(s) in the data to fit a regression model and make accurate predictions.

To improve upon the DT, RF and ET prediction accuracy, Regression Chain can be used.[39] This approach involves fitting a series of models to predict each output variable in a sequential manner. Taking the output variable sequence $G$, $N_{TR}$, $E_a$ as an example, firstly, a model ($F_1$) is fitted using the train-validation dataset 252 × $X(32)$ to predict the first output variable ($G$). Then, a second model ($F_2$) is fitted using a modified dataset, created by concatenating the base train-validation dataset and the actual values of the first output variable ($G$), to predict the second output variable ($N_{TR}$). Finally, in the last stage, model ($F_3$) is created using the third output variable ($E_a$) and concatenated data (train-validation dataset, $G$, $N_{TR}$).

During the validation process, predictions for the first output variable ($\hat{G}$) are made using model ($F_1$). Then, the prediction of $\hat{G}$ is added to the test dataset 28 × $X(32)$, and the second output variable ($\widehat{N_{TR}}$) is predicted using model ($F_2$). Finally, the first two predictions ($\hat{G}$ and $\widehat{N_{TR}}$) are concatenated with the test dataset, and the third output variable ($\widehat{E_a}$)) is predicted using model ($F_3$).

The main problem with this method is that the randomness in determining the chain sequence (the sequency of the three ODE parameter values--$G$, $N_{TR}$, $E_a$) can lead to significant differences in predictive performance. This randomness can impact the accuracy and reliability of the predictions. By sequentially predicting each output variable, the model can capture the relationships and dependencies among the variables, which can be valuable in certain scenarios. The order in which the different values of the ODE parameters are predicted is determined based on the loop sequence during simulation ($G$, $N_{TR}$, and $E_a$).

**ML Model evaluation**

The performance of the ML models is evaluated using correlation coefficient $R$, shown in equation 10 [40]:

$$R = \frac{\sum_{i=1}^{n}(y_i - \bar{y})(\hat{y}_i - \bar{\hat{y}})}{\sqrt{\sum_{i=1}^{n}(y_i - \bar{y})^2}\sqrt{\sum_{i=1}^{n}(\hat{y}_i - \bar{\hat{y}})^2}}, \qquad (10)$$

$R$ is used as a measure of the relationship between the true ODE parameter values $y_i$ and predicted ODE parameter values $\hat{y}_i$. $n$ is 252 and 28 in train-validation set and test set. $R$ can have a value between -1 and 1, where 1 represents a perfect positive linear correlation between the true ODE parameter values and predicted ODE parameter values, which indicates the good model prediction performance. 0 represents no relationship.

We also use the normalized root mean squared error ($NRMSE$), shown in equation 11:

$$NRMSE(\%) = 100 * \frac{\sqrt{\sum_{i=1}^{n}\frac{(y_i - \hat{y}_i)^2}{n}}}{y_{i_max} - y_{i_min}} \quad (11)$$

where $y_{i_max}$ and $y_{i_min}$ denoted the maximum and minimum ODE parameter values, respectively. $NRMSE$ is a commonly used measure to evaluate the performance of regression models. It is calculated by dividing the RMSE (Root Mean Squared Error) by the range of the ODE parameter values. The range is the difference between the maximum and minimum values of the ODE parameter values. $NRMSE$ provides a normalized measure of the ML model's error and allows for comparison of the ML model's performance across different datasets and ODE parameter values with different scales. $NRMSE$ values range between 0 and 100 %, with 0 indicating a perfect fit and 100 % indicating that the ML model's predictions are as accurate as the mean value of the ODE parameter values.

## 4 Results and discussion

### *4.1 model selection*

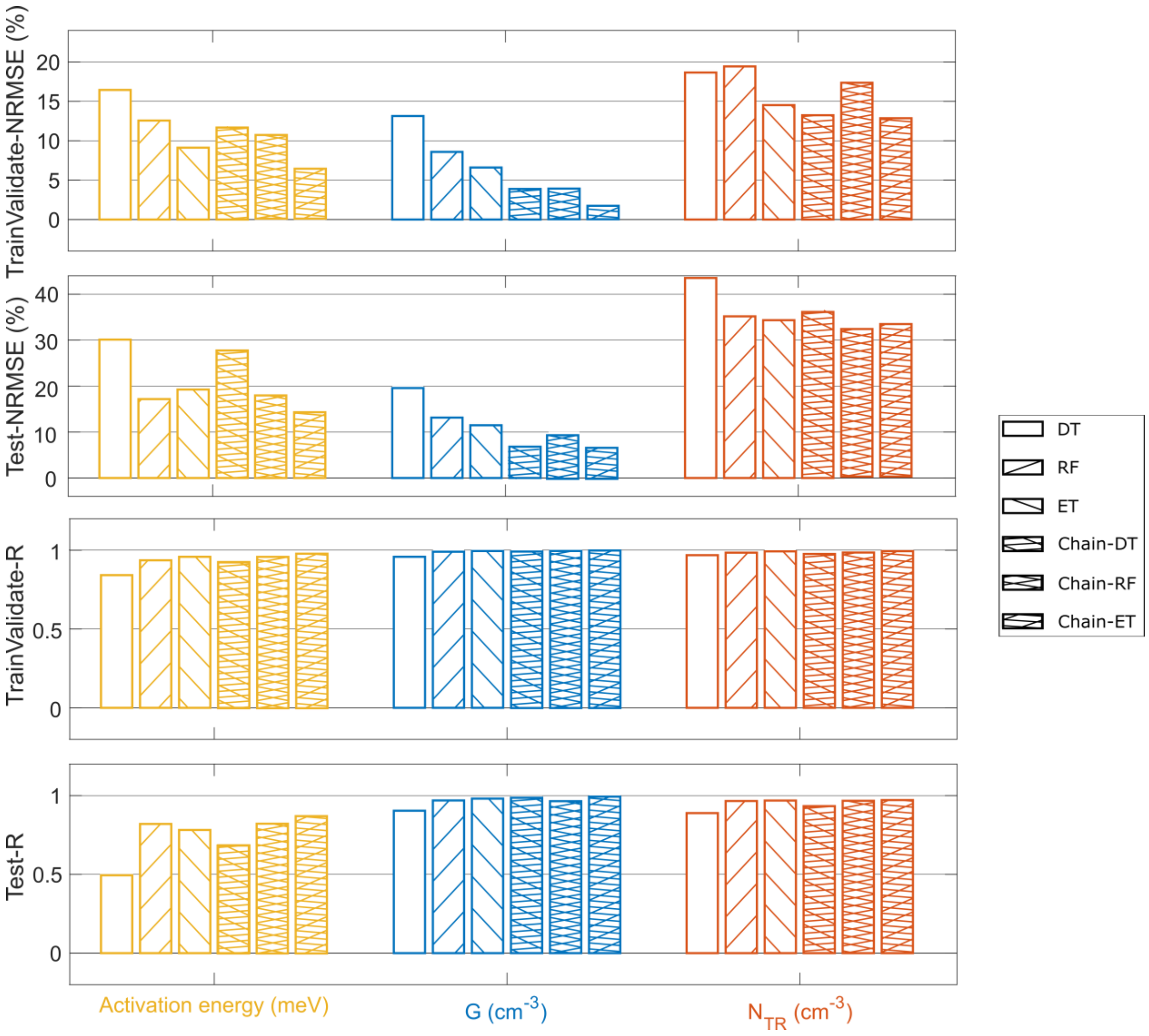

**Figure 4** Model performance comparison based on $NRMSE$ and $R$ for predicting the three ODE parameters using the train-validation dataset and test dataset. The top two rows show the $NRMSE$ values on the train-validation dataset and test dataset, respectively, while the bottom two rows show the $R$ values on the train-validation dataset and test dataset, respectively for each ML model: DT, RF, ET, Chain-DT, Chain-RF, Chain-ET. The columns from left to right shows the performance of the activation energy (marked with yellow color), exciton generation rate (marked with blue color), and total trap concentration (marked with orange color), respectively.

Figure 4 compares the performance of different models in predicting the three ODEs parameters based on $NRMSE$ and $R$ for both the train-validation dataset and test dataset. The results show that the prediction of $G$ has high $R$ and low $NRMSE$ for both datasets. Chain-ET has significantly improved the prediction performance compared with the other five ML models (DT, RF, ET, Chain-DT, and Chain-RF).

For $G$, the $R$ and $NRMSE$ of the train-validation dataset prediction are about 0.99 and 2%, respectively, and they are about 0.99 and 7% for the test dataset prediction. For $N_{Tr}$, the $R$ and $NRMSE$ of the training-validation dataset prediction are about 0.99 and 8%, respectively, while for the test dataset prediction, they are about 0.99 and 17%.

Predicting $N_{TR}$ and $E_a$ are challenging because they have opposite effects on $N_{Tr}$. For $E_a$, the $R$ and $NRMSE$ of the train-validation dataset prediction are about 0.98 and 6%, respectively, while for the test dataset prediction, they are about 0.87 and 14%. For $N_{TR}$, the $R$ and $NRMSE$ of the train-validation dataset prediction are about 0.99 and 13%, respectively, while for the test dataset prediction, they are about 0.97 and 33%.

*4.2 prediction using simulated data — test dataset*

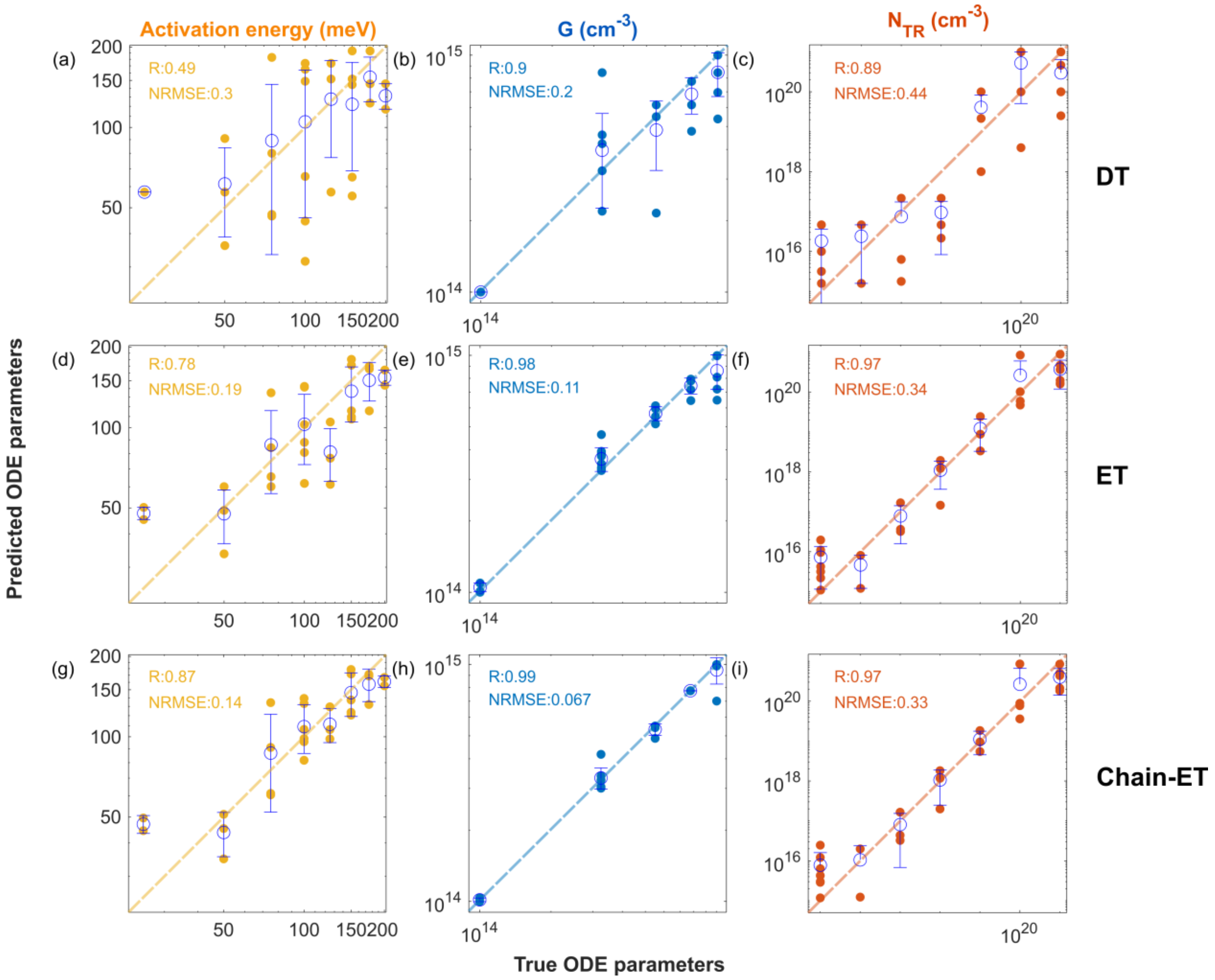

**Figure 5** Scatter plots of the prediction performance of the DT, ET, and Chain -ET for the three ODE parameters ((a, d, g) trap activation energy ($E_a$), (b, e, h) generation of exciton ($G$), and (c, f, i) total trap concentration ($N_{TR}$) on the test dataset. Standard deviations of 35, 56, and 40 repetitions for each value of the activation energy, generation of exciton, and total trap concentration, respectively, correspond to the error bars in the figure.

Figure 5 shows the scatter plots of the evaluation of the optimized DT, ET, and Chain-ET models on the simulated data--test dataset, which was not used during train-validation. The results demonstrate that the Chain-ET model performs well on the test dataset with high correlation coefficients (0.87, 0.99, 0.97) for the activation energy, generation of exciton ($G$), and total trap concentration ($N_{TR}$)), respectively. The Chain-ET model outperforms DT and ET models with a 25 % and 4 % improvement in average $R$ performance among three ODEs parameters, respectively. The high correlation coefficients on the test dataset indicate that the ML models can predict the three parameters accurately on unseen data, indicating their high predictive capability. The Chain-ET model shows a 42 % and 17 % improvement in average $NRMSE$ compared to the DT and ET models, respectively. The $NRMSE$ values for these three ODE parameters are 14%, 6%, and 33%, respectively. In addition, the standard deviation (STD) from the Chain-ET has improved as well, which indicate a less dispersion in relation to the predicted mean value. Therefore, the Chain-ET algorithm enhances the prediction accuracy of the ODE parameters on the unseen dataset. And the optimized Chain-ET method application on the temperature intensity modulation two photon spectroscopy was called Mlim (**M**achine **l**earning regression in **i**ntensity **m**odulation spectroscopy) in this work.

*4.3 prediction using experimental data*

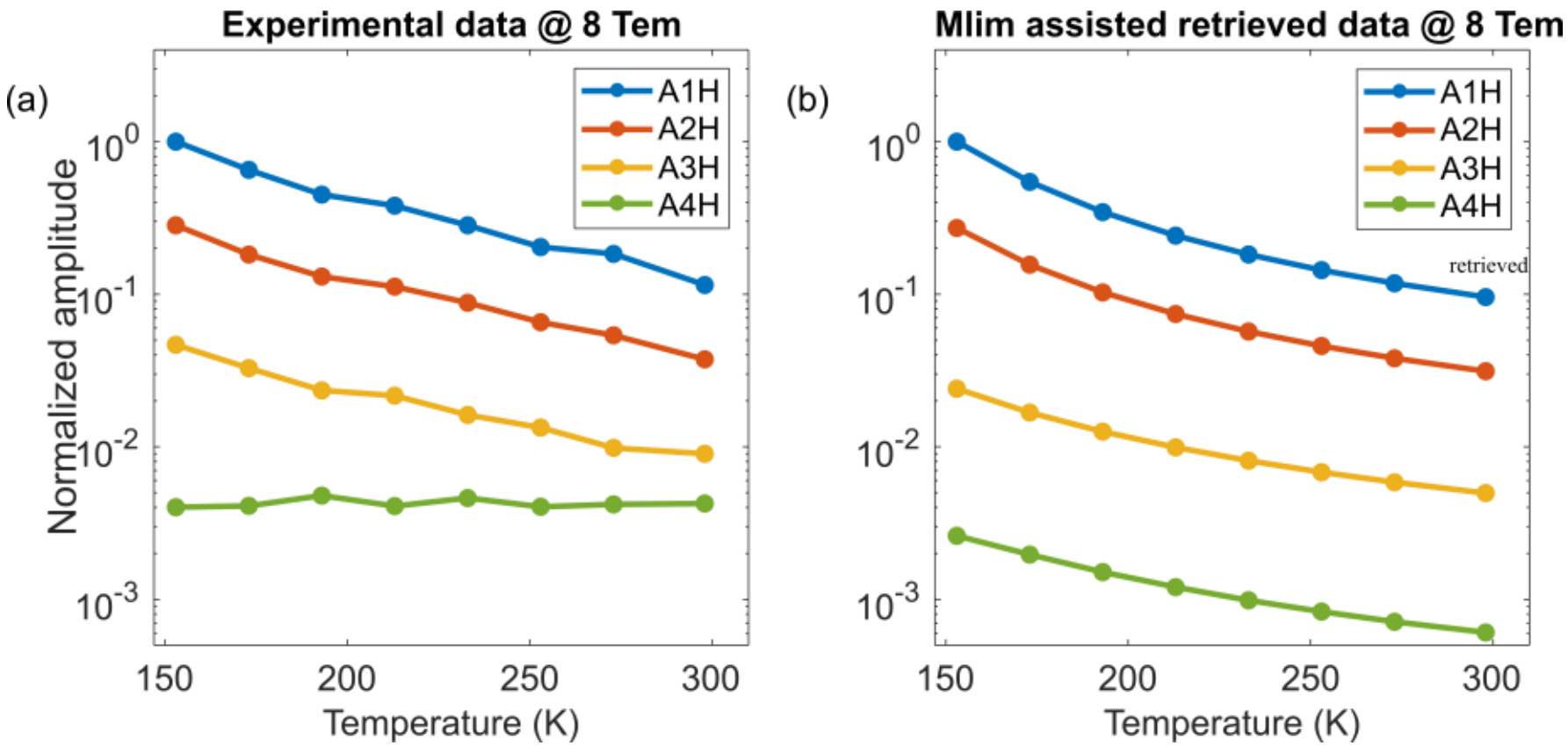

Figure 6. The experimental data used as input to the Chain-ET regression: Normalized modulated PL spectra (A1H, A2H, A3H, and A4H) of (a) experimental data and (b) Mlim assisted retrieved data at 8 temperatures (150, 170, 190, 210, 230, 250, 270, 300 K respectively).

Once the Chain-ET model has demonstrated good performance, it can be applied to predict the experiment results to reveal the photophysical characteristics of perovskite thin film samples. The IM2TM experimental data was gathered as part of previous work[4], with the specifics of the experimental setup have been described elsewhere,[6,41,42] and outlined in Section S4.

Figure 6 provides an overview of the results obtained when applying the Mlim model to the experimental data, focusing on $MAPbBr_3$ perovskite thin films. Figure 6a displays four harmonics PL emission (A1H, A2H, A3H, A4H) at eight different temperatures. The experimental data was used as the input dataset to the Mlim model (X (32)), which yielded three output datasets (Y (3)). The predicted value of generation of exciton ($G$), total trap concentration ($N_{TR}$), and trap activation energy ($E_a$) are as follows: $9.0 \times 10^{14}\ cm^{-3}$, $7.6 \times 10^{17}\ cm^{-3}$, 50 meV, respectively. And the activated trap concentration can be calculated accordingly to be about $1.1 \times 10^{17}\ cm^{-3}$ at room temperature. The schematics demonstrate the prediction of the key parameters in the charge carrier recombination dynamics by Mlim are shown in Figure 7 a-b.

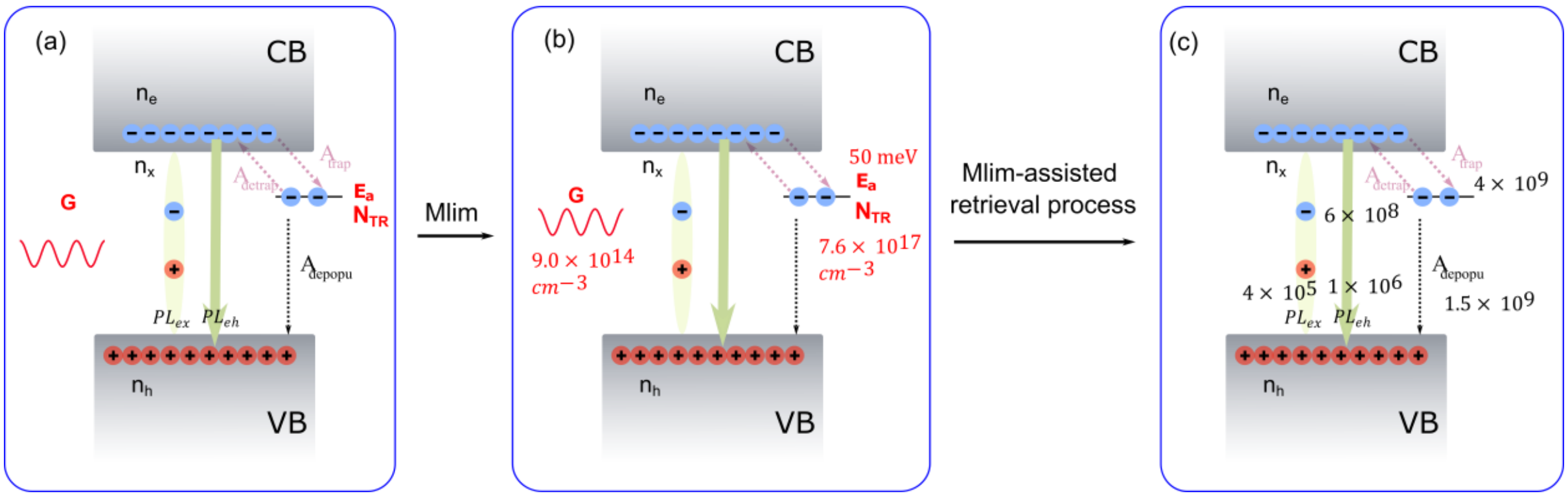

Figure 7. Illustrated schematics demonstrate the application of Mlim in unraveling the recombination mechanisms of perovskite microcrystal films: (a) A charge carrier recombination dynamics model is presented, featuring the conduction band (CB) and valence band (VB), denoting electrons in the CB ($n_e$- light blue color) and holes in the VB ($n_h$ - light red color). Excitation generation ($G$), activation

energy of traps ($E_a$), trap concentration ($N_{TR}$), PL emission from excitons ($PL_{ex}$), and PL emission from free electron-hole pairs ($PL_{eh}$) are showcased. The amplitudes of trapping, detrapping, and nonradiative depopulation are indicated as $A_{trap}$, $A_{detrap}$, and $A_{depopu}$, respectively. (b) Applying Mlim to experimental data of $MAPbBr_3$ perovskite microcrystals allows for the prediction of excitation generation ($G$), activation energy ($E_a$), and trap concentration ($N_{TR}$). (c) Through the Mlim-assisted retrieval process, the amplitudes of trapping ($A_{trap}$), detrapping ($A_{detrap}$), and nonradiative depopulation ($A_{depopu}$) are determined, completing the characterization of recombination mechanisms.

To validate the reliability of these three predicted ODE parameters, which provide insights into the trap properties within the perovskite, they were used to retrieve the temperature-dependent intensity modulation spectra and compared with experimental data. Figures 6 b depicts the retrieved temperature-dependent intensity modulation spectra using Mlim. The relative errors for A1H, A2H and A3H are small, while the larger mismatches observed in the higher harmonics (A4H) between the experimental and retrieved spectra can be attributed to multiple factors. First, it may originate from the fluctuation of the cooling panel in the temperature control system. Second, the charge accumulation effect, previously discussed in our earlier work[5], significantly influences the amplitudes and phases of the higher harmonic signals. However, the machine learning regression model in this work does not incorporate phase information. Moreover, the performance of Mlim relies heavily on the charge carrier recombination dynamics models proposed (equations 1-3) and the ODE parameters referenced in Table 1. There may be multiple types of traps and varying relaxation rate constants for different relaxation processes within such materials. All of these potential factors should be considered during the data engineering process and will serve as the ground for our future research. While there are mismatches between the retrieval results and the experimental spectra, the Pearson coefficient between them is 0.987.

Through the Mlim-assisted retrieval process, we gain insights into the behavior of charge carrier dynamics, unveiling the intricacies of relaxation processes. At room temperature, the amplitudes of various relaxation processes are as follows: excitons PL emission ($4 \times 10^5$), free electron-hole pairs PL emission ($1 \times 10^6$), trapping ($4 \times 10^9$), detrapping ($6 \times 10^8$), and depopulation processes ($1.5 \times 10^9$). The schematics demonstrate the retrieval process from the key parameters in the charge carrier recombination dynamics to unravel different relaxation processes assisted by Mlim are shown in Figure 7 b-c.

Despite current challenges, our charge carrier recombination models can still capture most of the key features of charge carrier recombination dynamics in $MAPbBr_3$ perovskite thin films. The optimized machine learning regression method, Chain-ET (Mlim), could help to reveal the detailed trap features, such as total trap concentration, trap activation energy, and activated trap concentration in $MAPbBr_3$ perovskite films. This can be further improved by retraining the regressor using more extensive training datasets and comprehensive data engineering, which includes phase information for each harmonic & temperature and more variations in charge carrier recombination dynamics parameters.

Lastly, by directly comparing the Mlim-retrieved intensity modulation spectra with experimental data, the reliability of trap property parameters is confirmed. Besides, the intricacies of the charge carrier recombination dynamics can be unraveled. In the future, this approach could be extended to temperature-dependent intensity modulation two-photon microscopy of perovskite thin films, enabling high-resolution mapping of trap states and their impact on charge carrier relaxation dynamics in perovskite microcrystals. This suggests that Mlim applications hold promise for studying photoactive devices.

## 5 Conclusion

In conclusion, to fully explore the potential of perovskite materials, we integrate functional intensity modulation two-photon spectroscopy with AI-enhanced data analysis to comprehensively explore defect-related trap states in perovskite microcrystals.

Firstly, we have introduced a charge carrier recombination dynamics model that takes into account both exciton and electron-hole pair photoluminescence (PL) emissions, as well as trapping-detrapping equilibrium. This model serves as a valuable approach for characterizing trap properties and understanding the photophysics in $MAPbBr_3$ perovskite microcrystals. By varying the parameters model (two-photon absorption, total trap concentration, trap activation energy) within the dynamic model, we generate a pool of temperature-dependent intensity modulation photoluminescence spectra by solving the ordinary differential equations in the charge carrier dynamics model. Subsequently, we employ tree-based supervised machine learning methods (Decision tree, Extra tree, and Random Forest) in conjunction with the ensemble technique of regression chain to optimize our **M**achine **L**earning **I**ntensity **M**odulation Spectroscopy (Mlim). Remarkably, the Chain-extra tree model achieved remarkable correlations exceeding 98% for the training-validation set and over 87% for the test set, with low NRMSE values below 13% for the training-validation set and 33% for the test set.

The optimized Chain-extra tree machine learning regressor was then applied to the experimental measurement of a $MAPbBr_3$ perovskite film. The predicted three key parameters were subsequently used to retrieve experimental data, and the reliability of these parameters was confirmed by comparing the experimental data with the data obtained using Mlim-assisted retrieval. Notably, the Pearson correlation coefficient was approximately 0.987, demonstrating the accuracy of the predictions.

In addition, the analyses allow to map out various charge carrier recombination processes, such as exciton PL emission, free electron-hole PL emission, trapping, detrapping, and SRH losses, providing a comprehensive understanding of perovskite material photophysics.

## Acknowledgments

We acknowledge financial support from Swedish Energy Agency and Swedish Research Council. Collaboration within NanoLund is acknowledged. The authors thank K. J. Karki and P. Kumar for valuable help with experiment. The data handing was enabled by resources provided by LUNARC. ChatGPT was used to polish parts of the text of the article. We thank Andreas Jakobsson for valuable discussion.

## Supporting Information

S1. Parameters used in the set of coupled ordinary differential equations.
S2. Hyperparameters optimization.
S3. Representation of decision tree.
S4. Optical setup.
S5. Abbreviations in this paper.

# Supporting Information

## Machine learning regression analyses of intensity modulation two-photon spectroscopy (Mlim) in perovskite microcrystals

Qi Shi[1]*, and Tönu Pullerits [1]*

[1] The Division of Chemical Physics and NanoLund, Lund University, Box 124, 22100 Lund, Sweden

AUTHOR INFORMATION

**Corresponding Authors**

* qi.shi@chemphys.lu.se

* Tonu.Pullerits@chemphys.lu.se

### S1. Parameters used in the set of coupled ordinary differential equations

Table S1: Fixed parameters used in the set of coupled ordinary differential equations for Figure 1.

| Parameter | Symbol | Value | References |
|---|---|---|---|
| Exciton formation rate from free electron and hole | $R_f$ | $10^{-12}\ cm^3 s^{-1}$ | 1,2 |
| Exciton dissociation rate to electron and hole | $R_d$ | $5 * 10^{11}\ s^{-1}$ | 3 |
| Trapping rate | $\gamma_{trap}$ | $8 * 10^{-9}\ cm^3 s^{-1}$ | 4 |
| Detrapping rate | $\gamma_{detrap}$ | $10^7\ s^{-1}$ | 4 |
| Electron hole radiative recombination rate | $R_{eh}$ | $5 * 10^{-11}\ cm^3 s^{-1}$ | 4 |
| Exciton radiative recombination rate | $R_X$ | $5 * 10^7\ s^{-1}$ | 5 |
| SRH nonradiative recombination rate | $\gamma_{depopu}$ | $5 * 10^{-9}\ cm^3 s^{-1}$ | 4 |
| Trap concentration | $N_{TR}$ | $10^{19}\ cm^{-3}$ | |
| Trap activation energy | $E_a$ | $150\ (meV)$ | 6 |
| Generation rate of excitons | $G$ | $10^{15}\ cm^{-3}$ | 7 |
| Activated trap concentration | $N_{Tr}$ | $N_{TR} \times e^{-\frac{E_a}{k_B T}}$ | |

### S2. Hyperparameters optimization

In this study, we performed hyperparametertuning for three different tree-based models: decision tree (DT), random forest (RF), and extra tree (ET). In decision trees, the hyperparameters that can be tuned include the maximum depth of the tree (max_depth), the minimum number of samples required to split an internal node (min_samples_split), the maximum number of features that are considered when splitting a node (max_features), and the minimum number of samples required to be at a leaf node (min_samples_leaf). In random forests, the number of trees in the forest (n_estimators), (max_depth), and the (min_samples_split) and (min_samples_leaf) are important hyperparameters. In extra trees, (n_estimators), the (min_samples_split) and (min_samples_leaf) hyperparameters that can be tuned. The optimized hyperparameters are marked with blue color and are shown in Table S3.

Table S2: Optimized architecture and hyperparameters obtained using the proposed optimization.

Hyperparameters for Random Forest

| Parameter | Value |
|---|---|
| max_depth | [30,50,70, 90, 110,**130**,150,180,200,300,500,700,900,1100,1300,1500,1700] |
| min_samples_leaf | [**2**, 3,4,5,6] |
| min_samples_split | [2, 3,**4**,5,6] |
| n_estimators | [5,10,25,**50**,100,150,200,250,300,350] |

Hyperparameters for Extra tree

| Parameter | Value |
|---|---|
| min_samples_leaf | [**2**, 3,4,5,6] |
| min_samples_split | [2, 3,**4**,5,6] |
| n_estimators | [5,10,25,50,**100**,150,200,250,300,350] |

Hyperparameters for Decision tree

| Parameter | Value |
|---|---|
| max_depth | [30,50,70, 90, 110,130,150,180,200,300,500,700,900,1100,1300,**1500**,1700] |
| min_samples_leaf | [2, **3**,4,5,6] |
| min_samples_split | [2,3,4,5,**6**,7,8,9,10] |
| max_features | [4,5,6,7,8,**9**,10] |

The optimial hyperparameters for each ML model are bolded.

## S3. Representation of decision tree

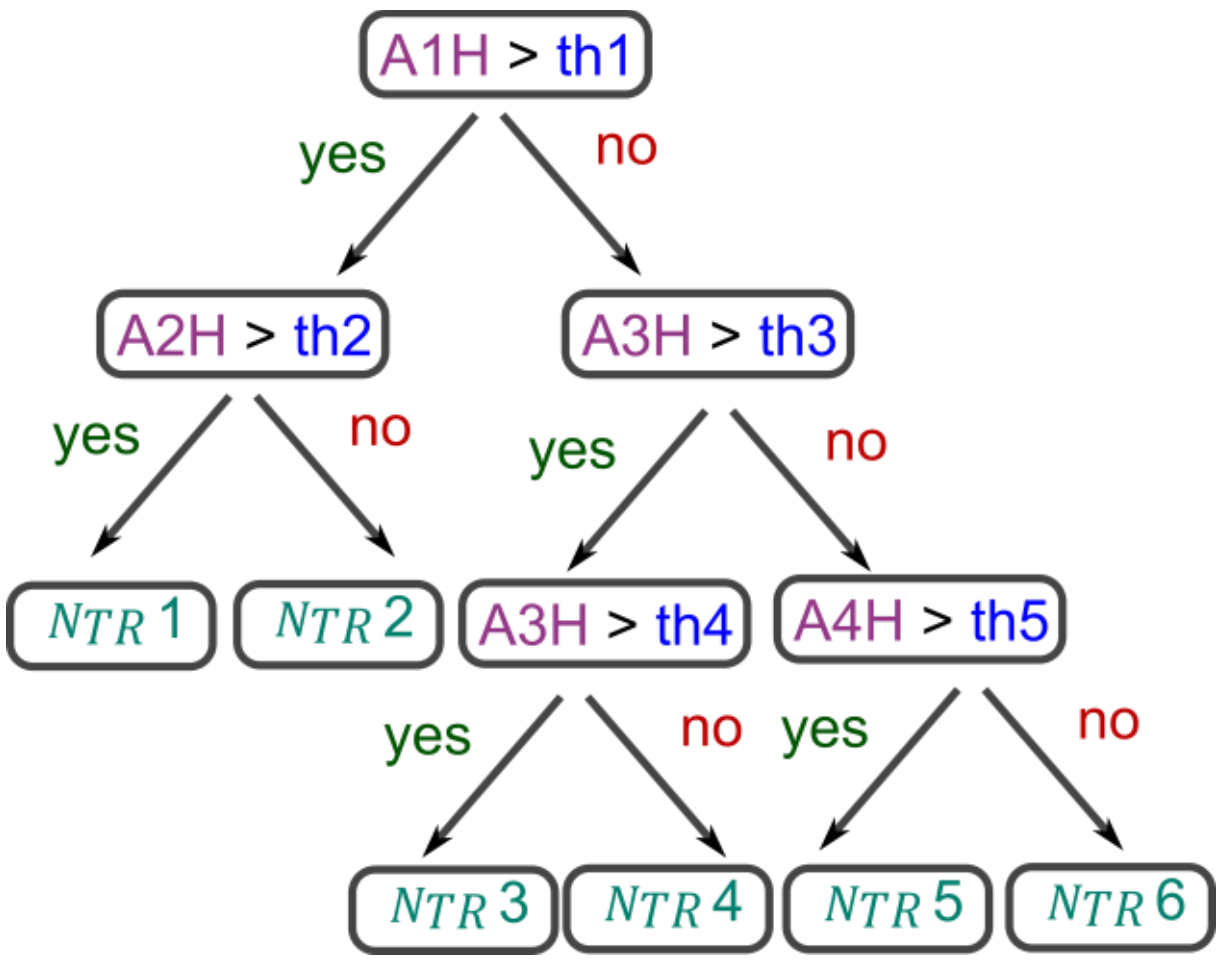

**Figure S1.** Representation of simple decision tree for prediction of total trap concentration ($N_{TR}i$ $(i = 1 \dots 6)$) based on the spectra features at room temperature ($AiH(T)_{i=1\dots4}^{T=298\,K}$). $Thi(i = 1 \dots 5)$ represents the threshold values used for if-then-else comparison. In this tree, $A1H$ is called 'root node', $AiH_{i=2,3,4}$ is called node. $N_{TR}i$ $(i = 1 \dots 6)$ is called 'leaf node'. The process of dividing a node into two or more sub-nodes is called 'splitting'.

## S4. Optical setup

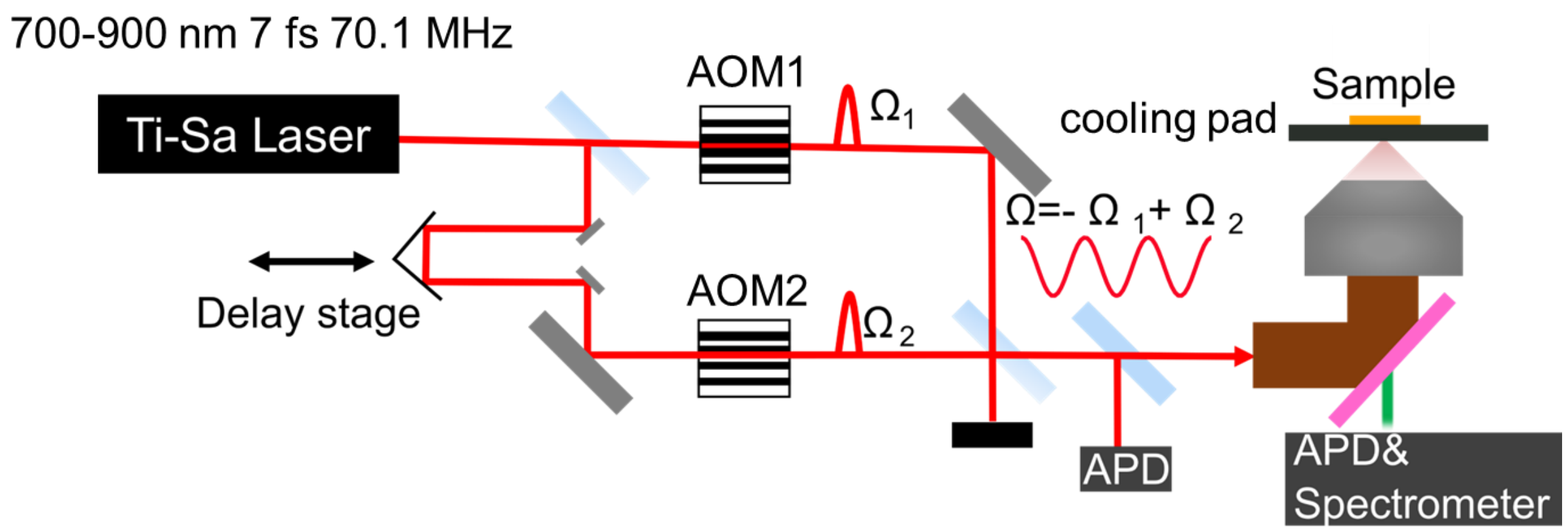

Figure S2 depicts the optical setup employed in this study. A Ti-Sapphire laser-based mode-locked oscillator serves as the light source, generating broadband laser pulses with a spectral range spanning from 700 nm to 900 nm and a pulse duration of approximately 7 fs. To counteract group velocity dispersion introduced by the optical components within the entire setup, a pair of chirped mirror pairs is employed to mitigate this effect.

In a Mach-Zehnder interferometer, a time-dependent phase change is introduced into each arm using two acousto-optic modulators (AOM) to modulate the average laser beam intensity. The AOMs operate at radio wave frequencies of 55 MHz and 54.95 MHz, with a frequency difference of 50 KHz. Following the third beam splitter, one portion is detected by an avalanche photodiode (APD) and served as a reference, enabling the tracking of laser intensity fluctuations, while the other portion is directed to an inverted microscope.

The microscope is equipped with a dichroic mirror that reflects light wavelengths longer than 650 nm and transmits shorter wavelengths. A reflective objective (RO) with a numerical aperture of 0.65 precisely focuses the laser beam onto the sample. Two-photon-induced photoluminescence (PL) intensity is detected by an avalanche photodiode (APD) with a bandwidth of approximately 5 MHz.

A temperature-controlled stage (Linkam Scientific Instruments, LTS420E-P) is employed, enabling temperature variations in the sample at 298, 273, 253, 233, 213, 193, 173, 153, 133, and 113 K. Please note that temperature values in this work are rounded for simplicity.

**S5. Abbreviations in this paper**

A1H First harmonic PL emission

A2H Second harmonic PL emission

A3H Third harmonic PL emission

A4H Fourth harmonic PL emission

CB Conduction band

DFT density functional theory

DT Decision tree

$E_a$ trap activation energy

ET Extra tree

FFT Fast Fourier transform

$G$ two-photon absorption

IM2PM Intensity modulation two-photon excited PL microscopy

LED light emitting diode

MAPE Mean absolute percentage error

ML machine learning

Mlim **M**achine **l**earning regression analyses of **i**ntensity **m**odulation two-photon spectroscopy

$NRMSE$ normalized root mean squared error

$N_{TR}$ total trap concentration

$N_{Tr}$ activated trap concentration

ODE ordinary differential equations

$P$ hyperparameter

PL photoluminescence

$R$ correlation coefficient

RF Random Forest

RT room temperature

SRH Shockley-Read-Hall

S-L Saha-Langmuir

VB Valence band